\documentclass[12pt]{amsart}

\newcommand{\R}{\mathbb R}
\newcommand{\x}{\mathbf x}
\newcommand{\y}{\mathbf y}
\newcommand{\kk}{\mathbf k}

\newcommand{\const}{\mathop{\rm const}\nolimits}

\begin{document}

\title[A condition for existence of $S$-matrix]
{A necessary condition for existence of $S$-matrix outside perturbation theory}
\author{A. V. Stoyanovsky}
\email{stoyan@mccme.ru}

\maketitle

\begin{abstract}
Using the Maslov--Shvedov method of complex germ, we show that quantum field
theory $S$-matrix can exist outside perturbation theory in the principal order
of quasiclassical approximation only under the condition that the tangent
symplectic transformation to the evolution operator of non-linear
classical field
equation is unitarily implementable in the Fock space. However, the results
of the book by Maslov--Shvedov imply that this condition is seemingly
always satisfied.
\end{abstract}

\section*{Introduction}

In this paper we study some consequences of Maslov--Shvedov's book [1], namely,
existence of the quantum field theory scattering matrix outside
perturbation theory in the principal order of quasiclassical approximation.
Our study is based
on a non-perturbative method of quasiclassical approximation developed by
V.~P.~Maslov and O.~Yu.~Shvedov and called the method of complex germ.
According to this method, in the principal order of quasiclassical approximation,
the evolution of a vector from the Fock space is given by the Bogolubov transformation
corresponding to the tangent symplectic transformation to the evolution
operator of classical non-linear field equations. Thus, in order to act as
a unitary operator in the Fock space, this symplectic operator should
satisfy the well known unitary implementability condition [2,3], namely, that
this symplectic operator be bounded in the $L_2$-norm, and certain operator built up
from the symplectic operator be a Hilbert--Schmidt operator.
It follows from the results of [1] that this condition is always satisfied.

The paper is organized as follows. In \S1 we recall construction of the $S$-matrix in
the Hamiltonian approach. In \S2 we recall the method of complex germ in Hamiltonian quantum
field theory. In \S3 we apply this method to evolution operator with the initial
conditions in the Fock space, and obtain the main result.

The author is deeply grateful to V.~V.~Dolotin and V.~P.~Maslov for
numerous illuminating discussions.

\section{$S$-matrix in the Hamiltonian approach}

For example, consider the model of scalar field in the spacetime $\R^{d+1}$ with the action
\begin{equation}
J=\int\left(\frac12\left(\left(\frac{\partial\varphi}{\partial t}\right)^2-(\nabla\varphi)^2
-{m^2}\varphi^2\right)-g(t)V_{int}(\varphi)\right)dtd\x,
\end{equation}
where $\x=(x_1,\ldots,x_d)$, and $g(t)$ is the interaction cutoff function.
The Hamiltonian has the form
\begin{equation}
\begin{aligned}{}
&H(t,\varphi(\cdot),\pi(\cdot))\\
&=\int\left(\frac12(\pi(\x)^2+(\nabla\varphi(\x))^2+m^2\varphi(\x)^2)
+g(t)V_{int}(\varphi(\x))\right)d\x.
\end{aligned}
\end{equation}
The Schrodinger equation for the wave functional $\Psi(t,\varphi(\cdot))$ reads
\begin{equation}
ih\frac{\partial\Psi}{\partial t}=\widehat H\Psi,
\end{equation}
where $\widehat H=H(t,\varphi(\cdot),-ih\frac{\delta}{\delta\varphi(\cdot)})$.
This equation is not well defined: in particular, $\frac{\partial^2\Psi}{\partial t^2}$
is undefined, because the square of the operator
$\widehat H$ has no sense even for $g\equiv0$.
To give a sense to equation (3),
one replaces the Hamiltonian $H$ by a regularized Hamiltonian $H_\Lambda$, where $\Lambda$ is
the regularization parameter, so that after removing regularization $\Lambda\to\infty$
the Hamiltonian $H_\Lambda$ goes to $H$, and for
$H_\Lambda$ instead of $H$, equation (3) is well defined. Then one adds to $H_\Lambda$
a summand $H^{ct}_\Lambda$ of order $O(h)$, called the counterterm Hamiltonian,
depending on fields $\varphi(\x)$, momenta $\pi(\x)$, the values of the function $g$ and its
derivatives at the point $t$, on the numbers $\Lambda$ and $h$. Then one considers the
equation
\begin{equation}
ih\frac{\partial\Psi}{\partial t}=(\widehat H_\Lambda+\widehat H^{ct}_\Lambda)\Psi.
\end{equation}
Note that
$H_\Lambda^{ct}$ is nonzero even for $g\equiv0$: in this case removing the regularization
yields the normally ordered Schrodinger equation for free field.

Further, one considers the equation with initial conditions in the Fock space of functionals
\begin{equation}
\Psi(\varphi(\cdot))=P(\varphi(\cdot))\Psi_0(\varphi(\cdot)),
\end{equation}
where
\begin{equation}
\Psi_0=\exp\left(-\frac1{2h}\int\tilde\varphi(\kk)\tilde\varphi(-\kk)\omega_\kk d\kk\right),
\end{equation}
\begin{equation}
\tilde\varphi(\kk)=\frac1{(2\pi)^{d/2}}\int e^{-i\kk\x}\varphi(\x)d\x,
\end{equation}
\begin{equation}
\omega_\kk=\sqrt{\kk^2+m^2},
\end{equation}
and $P(\varphi(\cdot))$ is a polynomial functional, or more precisely an element of the
Hilbert Fock space obtained by completion of polynomial functionals with respect to the
standard scalar product (see, for example, [1]). If one assumes that the function
$g(t)$ has compact support, and one develops the evolution operator of equation (4)
from $t=-\infty$ to $t=\infty$ into the series over the powers of the perturbation $g(t)$,
then in the so called renormalizable case, for appropriate $H^{ct}_\Lambda$, after
removing regularization we obtain a formal series $S(g)$ of operators in the Fock space,
called the Bogolubov $S$-matrix. To obtain the physical $S$-matrix, one makes the function
$g(t)$ tend to a constant $g\equiv\const$, called the interaction constant,
and one obtains a formal series with respect to this constant.

\section{The method of complex germ}

Consider equation (4) in the principal order of quasiclassical approximation, i.~e.,
up to $o(h)$. Denote by $H^{(1)}_\Lambda$ the principal term of the counterterm Hamiltonian,
so that $H^{ct}_\Lambda=hH^{(1)}_\Lambda+o(h)$. Maslov and Shvedov [1]
worked out a method of computation of the principal order of quasiclassical approximation
of the evolution of a vector of the Fock space, called the method of complex germ at a point.

To describe this method, introduce the following unitary operators in the Fock space:
\begin{equation}
P_{\Phi,\Pi,S}=\exp\left(\frac ihS-\int\Phi(\x)\frac{\delta}{\delta\varphi(\x)}d\x\right)
\exp\int\frac ih\Pi(\x)\varphi(\x)d\x,
\end{equation}
where $S$ is a real number, $\Phi(\x)$ and $\Pi(\x)$ are real functions from the Schwartz space.
Consider a vector of the form
\begin{equation}
\Psi(\varphi(\cdot))=P_{\Phi_0,\Pi_0,0}\cdot f_0\left(\frac{\varphi(\cdot)}{\sqrt h}\right).
\end{equation}
Then, according to [1], the result of the quasiclassical evolution of this vector from
the time $-T_0$ to the time $T$, after removing regularization, is the vector
\begin{equation}
\Psi(T,\varphi(\cdot))=P_{\Phi,\Pi,S}\cdot f\left(\frac{\varphi(\cdot)}{\sqrt h}\right),
\end{equation}
where  $\Phi=\Phi(T,\x)$, $\Pi=\Pi(T,\x)$ is the value at $t=T$ of the solution
of the classical equations of motion
\begin{equation}
\begin{aligned}{}
&\dot\Phi(t,\x)=\frac{\delta H}{\delta\pi(\x)}
=\Pi(t,\x),\\
&\dot\Pi(t,\x)=-\frac{\delta H}{\delta\varphi(\x)}
=(\Delta-m^2)\Phi(t,\x)-g(t)V_{int}'(\Phi(t,\x))
\end{aligned}
\end{equation}
with the initial conditions
\begin{equation}
\Phi(-T_0,\x)=\Phi_0(\x),\ \ \Pi(-T_0,\x)=\Pi_0(\x);
\end{equation}
$S$ is the action (1) on the classical trajectory $\varphi=\Phi(t,\x)$
from time $t=-T_0$ to $t=T$; $f=f(T,\xi(\cdot))$ is the result of evolution of the equation
\begin{equation}
i\frac{\partial f}{\partial t}=:H_2\left(t,\xi(\cdot),-i\frac{\delta}{\delta\xi(\cdot)}\right):f
\end{equation}
from $t=-T_0$ to $t=T$ with the initial condition $f(-T_0)=f_0$, where
\begin{equation}
H_2(t,\xi(\cdot),\eta(\cdot))=\frac12\xi\frac{\delta^2H}{\delta\varphi\delta\varphi}\xi+
\xi\frac{\delta^2H}{\delta\varphi\delta\pi}\eta
+\frac12\eta\frac{\delta^2H}{\delta\pi\delta\pi}\eta;
\end{equation}
here we write, for example, $\xi\frac{\delta^2H}{\delta\varphi\delta\varphi}\xi$ instead of
$$
\int\xi(\x)\frac{\delta^2H}{\delta\varphi(\x)\delta\varphi(\y)}(\Phi(t,\cdot),\Pi(t,\cdot))
\xi(\y)d\x d\y.
$$
In other words, $H_2$ is the quadratic part of the Hamiltonian $H$ at the point
$(\Phi(t,\cdot),\Pi(t,\cdot))$.

To be more precise and to explain the meaning of the signs $:\ldots:$ in formula (14),
assume that $f_0(\varphi(\cdot)/\sqrt h)$ is the vector (5) with $P(\varphi(\cdot))$ a
polynomial functional:
\begin{equation}
f_0\left(\frac{\varphi(\cdot)}{\sqrt h}\right)=M(u_1,v_1)\ldots M(u_n,v_n)\Psi_0,
\end{equation}
where
\begin{equation}
M(u(\cdot),v(\cdot))=
\int\left(u(\x)\frac{\varphi(\x)}{\sqrt h}-i\sqrt
hv(\x)\frac{\delta}{\delta\varphi(\x)}\right)d\x,
\end{equation}
for two functions $u(\x)$, $v(\x)$ from the Schwartz space. Then
\begin{equation}
\begin{aligned}{}
&f\left(t,\frac{\varphi(\cdot)}{\sqrt h}\right)=c(t)M(u_1^t,v_1^t)\ldots M(u_n^t,v_n^t)\\
&\times\exp\left(-\frac1{2h}\int\varphi(\x)R^t(\x,\y)\varphi(\y)d\x d\y\right),
\end{aligned}
\end{equation}
where the functions $u_i^t(\x),v_i^t(\x)$ satisfy the equations
\begin{equation}
\dot v^t_i(\x)=\frac{\delta H_2(v^t_i(\cdot),u^t_i(\cdot))}{\delta u_i^t(\x)},\ \
\dot u^t_i(\x)=-\frac{\delta H_2(v^t_i(\cdot),u^t_i(\cdot))}{\delta v_i^t(\x)}
\end{equation}
with the initial conditions $u^{-T_0}_i=u_i, v^{-T_0}_i=v_i$;
the operator $R^t$ with the kernel $R^t(\x,\y)$ satisfies the Riccati equation
\begin{equation}
\dot R^t+(R^t)^2-\Delta+m^2+g(t)V''_{int}(\Phi(t,\x))=0
\end{equation}
with the initial condition $R^{-T_0}=\sqrt{-\Delta+m^2}$.
Finally, $c(t)$ is a number depending on the counterterm
$H^{(1)}_\Lambda(\Phi(\cdot),\Pi(\cdot))$.
This counterterm is chosen so that this number be finite after removing regularization.

\section{Conditions for quasiclassical evolution in the Fock space}

Assume that the support of the function $g(t)$ lies between the points $-T_0$ and $T$.
Assume also that the evolution operator of equation (4) after removing regularization
is a well defined unitary operator in the Fock space (the $S$-matrix). Then the quasiclassical
evolution of the vector (10) for fixed solution $(\Phi(t,\x),\Pi(t,\x))$ of the classical
equations of motion should be also a well defined operator in the Fock space.
This evolution, given by equations (18--20), in particular, by the Riccati equation (20),
was studied in the book [1], in Sect.~8.3 of Chapter~6, independently of the assumption of existence of 
$S$-matrix. For example, 
the analysis of
{\it loc.~cit.}
implies that the result of evolution of the vacuum vector $\Psi_0$ (6) belongs to the Fock space.
Hence the same holds for the result of the evolution of the vector (16). 

Note that this evolution, defined on a dense subspace of the Fock space by the formulas above,
is exactly the well defined unitary transformation of the Fock
space corresponding to the tangent symplectic transformation to the Hamiltonian flow at the
point $(\Phi(t,\cdot),\Pi(t,\cdot))$ of the phase space of the field.
Hence this linear symplectic transformation
satisfies the well known unitary implementability condition [2,3],
namely, it is bounded in the $L_2$-norm,
and certain linear combination of its four components is a Hilbert--Schmidt operator.

Compare this result with the result of the paper [4] where it is shown that
the linear symplectic evolution operator of the free Klein--Gordon field between
curved spacelike Cauchy surfaces is not in general unitarily implementable in the Fock space.

\end{document}